\hsize=4.75 in
\hoffset=.75 in
\vsize=8 in
\voffset=.5 in
\font\large=cmr17

\noindent Published in {\it Quantum Classical Correspondence}: Proceedings
of the 4th Drexel Symposium on Quantum Nonintegrability, Philadelphia, PA,
USA, September 8-11, 1994, pp. 51-68.  Edited by Da Hsuan Feng and Bei Lok Hu.
Cambridge, MA: International Press, 1997.
\bigskip
\bigskip

\centerline{\large{True Collapse and False Collapse}}
\bigskip
\bigskip
\centerline{Philip Pearle}
\centerline{Hamilton College}
\centerline{Clinton, New York 13323}
\centerline{e-mail: ppearle@hamilton.edu}

\bigskip
\bigskip
\centerline{\bf Abstract}
\bigskip
	We emphasize that standard quantum theory (SQT) is incomplete because it
doesn't describe what is experimentally observed, namely events, nor does it satisfactorily
define the circumstances under which events may occur. 
Simple models are given (all of which 
have the same density matrix evolution) to illustrate schemes which
claim to complete SQT. It is
shown how the model based upon the Continuous Spontaneous Localization  (CSL) theory,
in which an individual statevector undergoes ``True Collapse," gives a satisfactory description of
events.  It is argued that various ``decoherence" based approaches, illustrated by models 
in which a  density matrix undergoes ``False
Collapse," do not satisfactorily resolve these problems of SQT.  
\bigskip
\bigskip
\noindent{\large{1 Introduction}}
\bigskip
{\it The abrupt change by measurement\dots is the most interesting point of the entire theory.}
\hfill Erwin Schr\"odinger$^{1}$
\medskip
{\it To restrict quantum mechanics to be exclusively
 about piddling laboratory operations is to betray the great enterprise.}
\hfill John Bell$^{2}$
\medskip
	What is wrong with standard quantum theory (SQT)? Doesn't it give wonderful agreement with 
every experiment so far performed? Then why should anyone wish to change it?

	I had better say what I mean by SQT.  I mean the usual formalism which results
in a statevector, taken to correspond to the physical reality we see around us
in Nature,$^{3}$ evolving with time, together with the Born probability rule. 
A structure which modifies (adds to, and/or alters) this shall either be called  
an interpretation of SQT or a 
different theory. 

	What is wrong with SQT is its description of quantum events. 
It doesn't describe them. Consider 
a particle counting experiment being conducted by an experimenter while a theoretician does a parallel 
real time calculation using SQT. The experimenter sets up the apparatus while the theoretician 
sets up the initial statevector. While the experimenter turns on the apparatus and monitors its 
smooth functioning, the theoretician follows the smooth evolution of the statevector
according to Schr\"odinger's equation.  Suddenly, the experimenter sings out ``An event
has occurred, and this is the result."  Abruptly, the theoretician stops his calculation, replaces
the statevector, which has by now become the sum of states corresponding to different possible 
outcomes of the experiment, by the one state which the experimenter told him had actually
occurred, and then continues his calculation of the smooth evolution of the statevector.

	In other words, the practitioner of
SQT must go {\it outside the theory}, to obtain additional
information, in order to use the theory correctly.  
What is missing is that the theory doesn't give the probability
that an event occurs between $t$ and $t+dt$. (One doesn't expect
the theory to give the precise time an event occurs:
that choice may be attributed to a fundamental probabilistic feature of nature.) Instead, the
Born probability rule says, {\it if}
an event occurs, {\it then} these are the possible results and their chances.  
Thus the description by SQT of {\it every} experiment is insufficient because 
events take place in every experiment and SQT fails to describe them.$^{4}$  

	A theory which gives a conditional probability alone 
is incapable of shedding light on the nature of the thing that is conditioned upon. Would 
we not say that something important is missing from a theory whose sole statement is, 
``If a cataclysmic event occurs in the solar system, then here are the possible things that can
 happen to the earth and their chances?" It might be useful to know the chances
of destruction and the forms it may take, but would we not be curious about 
the missing portion of the description?  Would we not want to know something about {\it when} 
it might occur, to inquire
about the causes of cataclysmic 
events, about what makes the sun explode, or an asteroid stray our way, even though
we might never be able to affect those events? 
How has it been that we have been so incurious about what makes quantum events?  Perhaps because
the situation has been more like, ``If a profitable financial event occurs, then here are the 
possibilities and their chances." We have been so busy  
reaping the profits of quantum theory that we have not cared to inquire about
the cause of the profitable events. 

	Tacit in this discussion was the assumption that
we know how to choose the``preferred basis" which specifies the catalog of 
possible events. But SQT doesn't tell
us this {\it either}.  Interpretations attempt to supply the missing information. 
The so-called (there are many variants) Copenhagen
interpretation (CI) 
 declares that the preferred basis describes the results of measurements
by an apparatus. But it doesn't say what is an apparatus or a
measurement (and it doesn't tell us if or how events can occur in
other situations).  Again, we must go outside the theory, to 
obtain ad hoc (``for this case only") information to use the theory. 
Useful? Wonderfully. CI is a  good practical guide
for working physicists. Measurement by an apparatus is
like great art, we know it when we see it---and we have great artists who can bring it about.  
But, a complete theoretical description?  No. 

	To summarize, SQT has two unsolved problems, which I shall call the ``events problem" (what
	is the probability that an event occurs in any time interval?)
and the "preferred basis" problem (given that the event occurs, what are the possible outcomes?),
i.e., ``when?" and ``what?"  	  

	How might one approach solving the events problem?  
Consider an analogous situation in classical mechanics.  An experimenter throws a ball 
toward a wall
while a theoretician follows its smooth evolution according to Newton's second law.  Suddenly
the experimenter sings out, ``It hit the wall, and this is now its position and velocity."  Abruptly
the theoretician stops the calculation, replaces the position and velocity by the new values and 
continues the calculation of the smooth evolution of the ball. Clearly one would think, ``This 
is not a complete theoretical description," and 
one would procede to incorporate the wall (and the ball's 
interaction with it) into Newton's second law so that the theoretical 
description is self contained. It seems equally clear that quantum theory needs
to have (what amounts to) the wall put in.

	In the Continuous Spontaneous Localization$^{4-6}$
(CSL) theory, a term is added to Schr\"odinger's equation,
resulting in a smoothly evolving statevector which describes events.  The 
modified equation produces

$$|\psi>\longrightarrow |a>\ \ {\rm OR}\ \ \ |b>\ \ \ {\rm OR}\ \ \ \dots \eqno(1.1)$$

\noindent instead of what SQT produces:

$$|\psi>\longrightarrow|a>+\ |b>+\dots \eqno(1.2)$$

\noindent The additional term 
contains a randomly fluctuating field, and it is 
this field which determines the events that actually occur.  Thus the theory
gives an explanation as to why we get a different result when we repeat the same 
experiment (the field fluctuated like {\it this} rather than like {\it that}), in contrast to
SQT which gives no reason at all why events occur.  

	Eq.(1.1) describes what I call True Collapse$^{7}$. 
In section 2, I \thinspace shall give a simple mathematical model of it, 
using the CSL formalism. This simple model does not do full justice to the physical ideas contained
in the full-blown CSL theory.  In particular, it does not show how the equations use 
the particle number basis to
solve the preferred basis problem, nor how they make collapse very slow for microscopic
objects and very fast for macroscopic objects.  
 (Also not touched upon here are interesting interpretational 
aspects of CSL, the pictures of spacetime 
reality that it allows, the successes and problems of a relativistic generalization,
the status of experimental tests: the reader is referred to the references.)
  
	There are a number of schemes, so-called  
``decoherence" interpretations, which claim to be able to describe events
without altering Schr\"odinger's evolution of the statevector, i.e., without providing
a collapse picture. In this paper I shall examine whether these schemes
do provide satisfactory criteria for identifying events.
In sections 3.1-3.3 \thinspace I shall give two simple examples
of one approach (dubbed here False Collapse of the first kind)   
 based upon the behavior of the density matrix 
describing an {\it ensemble} of statevectors, each of which evolves like (1.2). (Each 
statevector in the ensemble may be
thought of as describing the interaction of a system with
a different bath state.)
I shall argue that this gives no help in explaining 
the occurrence of events.

	In sections 4.1-4.2,\thinspace I shall give simple examples of
two schemes (dubbed here False Collapse of the second kind), the 
Environmental Selection Interpretation$^{8}$ (ESI) and the 
Decoherent Histories Interpretation$^{9}$ (DHI).
 They are based upon
the behavior of a partial trace of the 
density matrix constructed from a {\it single}  statevector which evolves like (1.2).
I shall argue that these, too, 
are not satisfactory.

	Lastly, in section 5, I shall
give a simple example of how a no-collapse scheme 
{\it can} account for events.  However, it requires adding new
dynamical variables to physics.   The point
I wish to make is that you cannot hope to describe events ``on the cheap,"
without {\it adding something new} to SQT.
\bigskip
\noindent{\large{2 A True Collapse Model}}
\bigskip
{\it One line of development would be to have the jump in the equations and not just the talk---
so that it would come about as a dynamical process in dynamically defined conditions. \dots
It would be good to know how this could be done.}

\hfill John Bell$^{10}$
\medskip
	For illustrative purposes I shall present a pale imitation of the full-blown
CSL theory.  Suppose that normal hamiltonian evolution has produced the statevector

$$|\psi ,0>=\alpha |a>+\beta |b>,\ \ \ \ |\alpha|^{2}+|\beta|^{2}=1\eqno(2.1)$$

\noindent where $|a>$ and $|b>$ are eigenstates of an operator $A$ 
with eigenvalues $a$ and $b$.  These states
may be thought of as describing different ``pointer positions," i.e., as macroscopically
different. The statevector evolution is presumed governed thereafter
by the modified Schr\"odinger equation

$${d|\psi,t>_{w}\over dt}=
-iH|\psi,t>_{w}-{1\over 4\lambda}[w(t)-2\lambda A]^{2}|\psi,t>_{w}\eqno(2.2)$$

\noindent with $H=0$ ($H$ may be supposed negligible because now it
only affects the internal evolution of the pointer states).  In Eq.(2.2), 
$w(t)$ is an arbitrary function of time (in the mathematical class of white noise ``functions"),  and 
$\lambda$ is the collapse rate (a fundamental parameter of the theory).  

	If you put any $w(t)$ into Eq.(2.2) you will find 
the statevector $|\psi,t>_{w}$ which has evolved under it.  But, what is
the probability that nature has chosen this $w(t)$?  To answer this question
the model supplies a second
equation, a probability rule:

$${\rm Prob}\{ w(t)\ {\rm for}\ 0\leq t\leq T\}=Dw\ _{w}
\negthinspace\negthinspace<\psi,t|\psi,t>_{w}\eqno(2.3)$$

\noindent Eq.(2.3) says that the largest
norm statevectors are most 
probable (the evolution (2.2) is nonunitary, so the statevector 
norm changes with time). 

	In Eq.(2.3), $Dw$ is the functional integral element:
	
$$Dw\equiv \prod_{t=0}^{t=T}{dw(t)\over \sqrt{2\pi\lambda (dt)^{-1}}}\eqno(2.4)$$
	
\noindent Since I shall use functional notation, I will say a bit about it. The way to 
look at an expression such as $\int_{-\infty}^{\infty}
Dw\exp -(2\lambda )^{-1}\int_{0}^{T} dtw^{2}(t)$ is 
to discretize it. That is, replace the continuous variable $t$ by $t_{n}$
replace the integral over $t$ by a sum over $n$, treat each $w(t_{n})$
like an ordinary variable and treat $dt$
like a number: thus this expression = 1.

	We can easily write the solution of Eq. (2.2) with the initial condition (2.1)
because the result only depends upon $B(T)\equiv\int_{0}^{T}dtw(t)$:
	
$$\eqalignno{|\psi,T>_{w}&=F\{ w(t)\}\cr
&\cdot\Bigl\lbrace \alpha |a>e^{-{1\over 4\lambda T}[B(T)-2\lambda Ta]^{2}}
+ \beta |b>e^{-{1\over 4\lambda T}[B(T)-2\lambda Tb]^{2}}\Bigr\rbrace&(2.5)\cr}$$

\noindent apart from an overall factor $F$ which is a functional I shall not bother to write down.
We can now use Eq.(2.3) to find the probability that a particular $B(t)$ 
(and the statevector it engenders)
occurs in Nature.  Taking the scalar product of (2.5)
with itself, and integrating over $w(t)$ for all $t$ except $T$, we obtain for the probability that
$B(T)$ lies in the interval $B$ to $B+dB$:

$${\rm Prob}={dB\over\sqrt{2\pi\lambda T}} \Bigl\lbrace|\alpha|^{2}e^{-{1\over 2\lambda T}[B(T)-2\lambda Ta]^{2}}
+|\beta|^{2}e^{-{1\over 2\lambda T}[B(T)-2\lambda Tb]^{2}}\Bigr\rbrace \eqno(2.6)$$

	From Eq.(2.6) we see that, as $T$ increases, the only probable values of
$B$ lie within a few standard deviations $\sqrt{\lambda T}$ of either $2\lambda Ta$ or $2\lambda Tb$.
In the former case, the statevector (2.5) asymptotically approaches const.$|a>$, while in the 
latter case it asymptotically approaches const.$|b>$. 
(The constants are of no significance since it is the statevector's direction in Hilbert space
that contains the information about reality: the statevector can
be renormalized at any time.)  Moreover, in the former case, the integrated probability (2.6)
asymptotically approaches $|\alpha |^{2}$ and in the later case approaches $|\beta |^{2}$.  Thus the 
statevector evolves into one or another of the macroscopically distinct states, and does so 
with the frequencies predicted by SQT.
	
	To summarize, Schr\"odinger's
evolution of the statevector can be modified, 
in a well-defined and reasonably simple way,
so as to give good agreement with SQT's predictions while describing events 
(paralleling the occurrence of each event in Nature
with collapse of a probable statevector). 

	We close this section by displaying the density matrix which describes the behavior of
the ensemble of (normalized) solutions to Eq.(2.2) which occur with probability (2.3):

$$\eqalignno{\rho (T)&=\int Dw\ _{w}\negthinspace\negthinspace
<\psi ,T|\psi ,T>_{w}{|\psi ,T>_{w}\null_{w}\negthinspace\negthinspace<\psi ,T|
\over _{w}\negthinspace\negthinspace<\psi ,T|\psi ,T>_{w}}\cr 
&=\int Dw|\psi ,T>_{w}\null_{w}\negthinspace\negthinspace<\psi ,T|\cr
&=\int_{-\infty}^{\infty}{dB\over \sqrt {2\pi\lambda T}}\Bigl\{
|\alpha |^2e^{-{1\over 2\lambda T}[B(T)-2\lambda Ta]^{2}}|a><a|&(2.7)\cr
 &\qquad\qquad\qquad\qquad+|\beta |^2e^{-{1\over 2\lambda T}[B(T)-2\lambda Tb]^{2}}|b><b|\cr
 &+\Bigl[\alpha \beta^{*}|a><b|
+\alpha^{*}\beta |b><a|\Bigr]e^{-{1\over 4\lambda T}[B(T)-2\lambda Ta]^{2}}
e^{-{1\over 4\lambda T}[B(T)-2\lambda Tb]^{2}}\Bigr\}\cr 
&=|\alpha |^2|a><a|+|\beta |^2|b><b|\cr&\qquad+\Bigl[\alpha \beta^{*}|a><b|
+\alpha^{*}\beta|b><a|\Bigr]e^{-{\lambda T\over 2}(a-b)^{2}}\cr}$$

\noindent  The reason
why the off-diagonal elements of the density matrix (2.7) are 
exponentially decaying is because the individual statevector 
collapses (corresponding to different $B(T)$'s)
occur at different times.
\bigskip
\noindent{\large{3.1 False Collapse Of The First Kind}}
\bigskip
{\it If it looks like a duck but it don't walk like a duck or quack like a duck
then it ain't a duck.}
\hfill Variant of an old saying
\medskip
	
  For my first example of a decoherence scheme I shall consider a  
a simple model of
the interaction of the system described 
by the operator $A$ with an environment. We suppose that
each time the system is put into a thermal bath it encounters a different bath state, 
and one can assign probabilities to these encounters. The claim is that 
interaction of the system with the bath brings on occurrences of events.

Consider the initial statevector (2.1) which evolves unitarily under the Schr\"odinger equation

$${d|\psi,t>_{w}\over dt}= -iw(t)A|\psi,t>_{w}\eqno(3.1)$$

\noindent In Eq.(3.1), $w(t)$ is a sample function of 
white noise (representing the influence of the bath).  That is, we assume 
repeated evolutions actually differ in  
w(t), whose probability of appearance satisfies the probability rule:

$${\rm Prob}\{ w(t)\ {\rm for}\ 0\leq t\leq T\}=Dwe^{-{1\over 2\lambda}\int_{0}^{T}dtw^{2}(t)}\eqno(3.2)$$

\noindent It follows from Eqs.(2.1) and (3.1) that the 
statevector at any time $T$ which evolves under a particular $w(t)$ is

$$ |\psi,T>_{w}=\alpha e^{-iB(T)a}|a>+\beta e^{-iB(T)b}|b> \eqno(3.3)$$

	Now, according to Eq.(3.3), NOTHING has HAPPENED, i.e., 
the state\-vector is still a superposition of
the states $|a>$ and $|b>$ with unchanged squared amplitudes. Of course, 
the phases of the amplitudes
are changing, but certainly one cannot find in this statevector evidence that
an event occurred or, supposing that an event occurred, whether 
it resulted in $a$ or $b$.

	Nonetheless, proponents of the point of view I am expounding are unphased (sic),
and they turn to the density matrix:

$$\eqalign{\rho (T)&=\int Dwe^{-{1\over 2\lambda}\int_{0}^{T}dtw^{2}(t)}|\psi ,T>_{w}\null
_{w}\negthinspace\negthinspace<\psi ,T|\cr
&=\int_{-\infty}^{\infty}{dB\over \sqrt {2\pi\lambda T}} e^{-{1\over 2\lambda T}B^{2}(T)}\Bigl\{
|\alpha |^2|a><a|+|\beta |^2|b><b|\cr &\qquad +\Bigl[\alpha \beta^{*}e^{-iB(T)(a-b)}|a><b|
+\alpha^{*}\beta e^{iB(T)(a-b)}|b><a|\Bigr]\Bigr\}\cr  
&=|\alpha |^2|a><a|+|\beta |^2|b><b|\cr &\qquad+\Bigl[\alpha \beta^{*}|a><b|
+\alpha^{*}\beta|b><a|\Bigr]e^{-{\lambda T\over 2}(a-b)^{2}}\cr}\eqno(3.4)$$

\noindent 
The False Collapse claim is that, at some large time, 
when the off-diagonal density matrix elements
are suitably small (in current parlance, at the decoherence time), an event 
($a$ or $b$) occurs for any system. 

This claim makes no sense, for two reasons.  
The behavior of an
{\it ensemble} of evolutions which have {\it not} taken place cannot be crucial in
determining the occurrence of an
event in {\it one} evolution which {\it did} take place (what is not real cannot have an effect
upon what is real). If no
individual statevector describes events, an ensemble of these statevectors cannot do so (a 
property missing in each element of a collection must be missing in the collection). Therefore
this scheme cannot solve the events problem. Although the density matrices (2.7) and (3.4) have
 the same form, this does not mean that the arguments leading to these expressions
are equally sound.
\bigskip
\noindent{\large{3.2 Against Decoherence}}
\bigskip
	In order to properly assess the meaning of Eq.(3.4), I believe it is salutary
to avoid using the phrase ``decoherence" time because that seems to imply that there is
a physical process called decoherence which takes this amount of time to be completed.  
I suggest that the phrase``No One Will Ever kNow" time, or NOWEN time for short, is more
apt for the following reason. 

	If the statevector evolution is as described in the previous section,    
then an event does not occur at any time. Suppose, however,
 you (wrongly) claim that an event occurs at time $T_{e}$.
 Suppose someone sets out to test your claim by performing a rapid experiment at time $T=T_{e}$
 ({\it then} causing an event to occur) which tests if the system is
in the state $|\phi >\equiv \mu |a>+\nu |b>$.  If an event ($a$ or $b$) did occur at time $T_{e}$, 
the system will be found to be in the state $|\phi >$ with probability 
$|\alpha|^{2}|\mu |^{2}+|\beta|^{2}|\nu |^{2}$.
But actually, it follows from Eq.(3.4) that the probability is

$$Tr\bigl\{|\phi ><\phi|\rho (T)\bigr\}
=|\alpha|^{2}|\mu |^{2}+|\beta|^{2}|\nu |^{2}+2\Re \{\alpha\beta^{*}\nu \mu^{*}\}
e^{-{\lambda T\over 2}(a-b)^{2}}\eqno(3.5)$$

\noindent The NOWEN time is, by definition,
a time at which the last term in Eq.(3.5) is small enough to be 
beyond experimental resolution.  Only if you chose $T_e$ to be equal to or greater 
than the NOWEN time will the experimental 
result be the same as if your claim were right.
With this choice, No One Will Ever kNow that you were wrong.
	   
\bigskip
\noindent{\large{3.3 A Simple Environment Model}}
\bigskip
	 The example in section 3.1 described the bath by the c-number w(t). Perhaps a more realistic 
example would be to include the bath states in the statevector.  In this case,
first, the density matrix for the ensemble of 
joint system + bath states is constructed.  Then the bath states are traced over, and the NOWEN
time is declared to be the time by which events have taken place.

	I shall give an example of this here, largely because the simple environment bath model 
will be useful in the remainder of this paper.

	Imagine a system which is bombarded by a succession of particles which briefly
interact with it and then travel away.  A simple model
which captures the essence is to suppose that the $n^{\underline{\rm th}}$ particle has
position coordinate $W_{n}$, and that it interacts with the system during the
time interval $(t_{n},t_{n}+\Delta t)$ with hamiltonian $AW_{n}$. Proceeding
to the limit $\Delta t\rightarrow 0$ of a continuum of operators 
$W_{n}\rightarrow W(t)$ which mutually commute,
we have the evolution equation 

$${d|\psi,t>\over dt}=-iAW(t)|\psi,t>\eqno(3.6)$$

	We shall denote a joint position eigenvector by $|\{w\}>$, i.e.,
$W(t)|\{w\}>=w(t)|\{w\}>:-\infty<w(t)<\infty$. (One may think of $|\{w\}>$ as the direct 
product of position eigenstates of $W(t)$ for each vaue of $t$.) 
We now construct a unity-normalized state $|\{w\}>_{N}$ 
 which is as close to an eigenstate of $W(t)$ with eigenvalue $w(t)$ as one wishes, i.e.,
 
$$|\{w\}>_{N}\equiv\prod_{t}\lbrace\int_{-\infty}^{\infty} dw'(t)
f[w'(t)-w(t)]\rbrace|\{w'\}>$$
 
\noindent where $f[w'-w]$ is sharply peaked 
at $w'-w=0$, and $\int_{-\infty}^{\infty} dw'f^{2}[w'-w]=1$. 
(f is essentially the square root of a delta function.)
 
	We suppose that, each time the system is put into
the bath of these bombarding particles, the initial statevector is

$$|\psi ,0>=\Bigl[\alpha |a>+\beta |b>\Bigr]|\{w\}>_{N}\eqno(3.7)$$

\noindent and the probability that the bath state $|\{w\}>_{N}$ appears with 
eigenvalues $\{w(t)\}$
is given by the white noise probability expression (3.2).

	The solution of Schr\"odinger's equation (3.6) with initial statevector (3.7)~is
	
$$ |\psi ,T>=\Bigl\lbrace\alpha e^{-iB(T)a}|a>+
\beta e^{-iB(T)b}|b>\Bigr\rbrace|\{w\}>_{N}\eqno(3.8)$$

\noindent As in section 3.1 we cannot infer that an event has 
occurred from such a statevector. (This statement
would still be correct had we chosen an initial statevector for the environment 
where f was not sharply peaked, so that each final state of system + environment would be an 
entangled one---see the next section.)  Also, it follows
from Eq.(3.8) that, if we construct the density matrix and trace over the environment states, 
we get the density matrix (3.4). This behavior of the statevector or density
matrix is in no essential way different
from what has already been described, and the critical remarks already expressed apply to it. 
\bigskip
\noindent{\large{4.1 False Collapse of the Second Kind}}
\bigskip 
 	{\it The idea that elimination of coherence, in one way or another, implies the
replacement of ``and" by ``or" is a very common one among solvers of the ``measurement problem."
It has always puzzled me.}
\hfill John Bell$^{2}$
\medskip
	In this section I shall address
the ESI.$^{8}$ 
Here, as in CSL, only one statevector represents the world.
However, the  world is divided into two:
 ``the environment" and the rest. 
But ESI does not define what is meant by environment any 
better than CI defines apparatus. This already is a basis (pun!) for
serious criticism: the very nature of the real world, the 
preferred basis, depends upon this undefined choice. 
Thus ESI does not resolve the preferred basis problem any better than does CI.
	
	Moreover, the choice of the division affects the 
	predicted times of events as well, so already the events problem is faring badly. But,
	never mind, suppose
that we somehow know what is the system and what is the environment. 
Consider the following initial 
statevector belonging to the system + environment modeled in section 3.3:
 
$$|\psi ,0>=\Bigl[\alpha |a>+\beta |b>\Bigr]\int 
D_{0}^{\infty}we^{-{1\over 4\lambda}\int_{0}^{\infty}dt w^{2}(t)}|\{w\}>\eqno(4.1)$$ 
	
\noindent where $D_{t_{1}}^{t_{2}}w\equiv \prod_{t 
=t_{1}}^{t_{2}}dw(t)/[2\pi \lambda /dt]^{1/4}$.  That is, each
particle's position has an initial gaussian probability distribution. 
The evolution equation is (3.6).  The 
resulting statevector is

$$ \eqalign{|\psi ,T>&=\Bigl\lbrace\alpha|a>\int D_{0}^{T}we^{-\int_{0}^{T}dt
w(t)[ia+(4\lambda )^{-1}w(t)]}|\{w\}>_{0}^{T}\cr 
&\qquad+ \beta |b>\int D_{0}^{T}we^{-\int_{0}^{T}dt
w(t)[ib+(4\lambda )^{-1}w(t)]}|\{w\}>_{0}^{T}\Bigr\rbrace\cr 
&\qquad\qquad\cdot \int D_{T^{+}}^{\infty}we^{-{1\over 4\lambda}\int_{T^{+}}^{\infty}dt
w^{2}(t)}|\{w\}>_{T^{+}}^{\infty}\cr }\eqno(4.2)$$

\noindent where $|\{w\}>_{t_{1}}^{t_{2}}$ is the joint position eigenstate of $W(t)$ for 
$t$ in the range $\{t_{1},t_{2}\}$.

	The statevector (4.2) shows that the states $|a>$ and $|b>$ are now entangled with
the ``environment" particle states labeled by $t$ from $0$ to $T$ (the particles labeled
by $t$ from $T^{+}$ to ${\infty}$ have yet to interact with the system, so their 
states are not yet entangled.)  But, just as with the statevectors (3.3) and (3.8), 
this statevector gives no indication that an event ($a$ or $b$) has occurred.  

	However, if one constructs the density matrix and traces over the environment, one gets
for this reduced density matrix precisely the last two expressions in (3.4). 
Again, the claim is made that either event $a$ or $b$ has taken place
by the NOWEN time, i.e., there
will be no difference between future behavior predicted by
using the uncollapsed statevector (4.2) or by using the mixture of collapsed statevectors 
$|a>|{\rm env}_{a},T>$ and $|b>|{\rm env}_{b},T>$ (present 
in respective amounts $|\alpha|^{2}$ and $|\beta|^{2}$).

	This ESI claim is demonstrably false.
An experiment could 
detect interference, which would be missing for the mixture,
 between the two system + environment states
which compose the actual statevector (4.2).

	Here is an example in the context of our model. 
Suppose each $W(t)$ represents the $x$-coordinate of 
an extremely massive particle at rest in the $z=0$ plane, with $y$-coordinate $t$ 
(take $0\leq t \leq T\geq$NOWEN time). Let
$A$ be the $x$-coordinate of 
an extremely massive particle (the A-particle) moving
in the $z=0$ plane in the $y$-direction with
constant speed 1. As the A-particle passes each environment particle $W(t)$ at time $t$,  
it briefly interacts with it. (The hamiltonian in the evolution equation (3.6) is consistent with
this picture). Let the initial state be (4.1) (with $D_{0}^{\infty}$ replaced by $D_{0}^{T}$). 
The A-particle is therefore in a superposition of two packets, one at $A=a$, the other at $A=b$.
 
	After the A-particle has passed the last environment particle at $y=T$ (resulting in the 
statevector (4.2) without the last line),
let it encounter a corner reflector (corner placed on the $y-$axis, with walls at $45^{\circ}$ to
the $y-$axis). Thus, when the A-particle has bounced off the two  walls, it has
changed the sign of its $x$-coordinate (so $|a>$ and $|b>$ have become $|-a>$ and $|-b>$) and 
reversed its direction of motion,
returning with speed 1 to collide with the row of environment particles in reverse order.

	It really isn't necessary,
but one can imagine
placing four mirrors which take the A-particle out of the $z=0$ plane for a while, and
bring it back so that it is once again headed to encounter the row of environment particles
in the order in which it first encountered them. Then, the Schr\"odinger evolution will effectively
be the time-reverse of the previous evolution.  That is, $-a$ ($-b$) is the eigenvalue of $A$ 
acting upon $|-a>$ ($|-b>$) as well as the eigenvalue of 
$-A$ acting upon $|a>$ ($|b>$).  Thus the effect of the evolution (3.6) upon the numerical coefficients 
in the state $\{$(4.2) with $|a>$ and $|b>$ replaced by $|-a>$ and $|-b>\}$
is the same as if $A$ were 
replaced by $-A$ and $|a>$ and $|b>$ were unchanged.

 Thus, after the A-particle's second trip past
 the environment particles, {\it the entanglement has been undone}. The resulting state's 
system part, 
$\alpha |-a>+\beta |-b>$, can certainly be distinguished from the purported
mixture of states $|-a>$ and $|-b>$.

	ESI proponents reply to this that what they mean by an environment is 
something so complicated that FAPP$^{11}$ it would be impossible to do such interference
experiments.  This is the NOWEN philosophy par excellence.  To base a
fundamental physical principle upon it is no more justified than basing morality upon the
principle that it is all right to steal if you don't get caught.
	
	We conclude that ESI doesn't solve either the preferred basis problem or the events problem. 
	
\bigskip
\noindent{\large{4.2 Against Decoherent Histories}}
\bigskip 
 	{\it It ain't over till it's over.}
\hfill Yogi Berra
\medskip
	
	I shall describe the DHI$^{9}$ 
 somewhat differently than usual, namely, that the evolution is that
of a single unitarily evolving 
statevector representing our world. (Usually, 
 a unitarily evolving density matrix is considered. If it is pure, it is equivalent
 to a single statevector. If it is impure, it is equivalent
 to an ensemble of statevectors, which was
 criticized in section 3.1.)
A certain ``decoherence condition," (see Eq.(4.3) et.seq.) is
applied to the density matrix constructed from this statevector.
 It is claimed that one can thereby extract 
 a set of ``histories," with each 
history characterised by a string of events (and a probability that
this string is the one realized in our world).
 For each history, we are told of the times of 
the events and projection operators associated with these times.  The predictions
are the same as for a True Collapse theory where instantaneous collapses are
characterized by applying each projection operator at the associated time to
the otherwise unitarily evolving statevector.
  
	Are a unique set of projection operators obtained from DHI? No.
As in ESI, the world is
divided into two parts, the part that is ``coarse grained" 
 and the part that is not (these may vary with time). Coarse graining 
  means that the variables of this part are 
 not to be distinguished by the projection operators. 
But, to be undistinguished does not mean to be unimportant.
 The coarse grained variables are usually crucial in determining which events are said to 
occur and when the events occur.
The ``environment" is a favorite candidate for these coarse grained variables.  But, just as in ESI, 
these variables are not specified by DHI, and so the preferred basis
problem is unsolved, and the events problem is given no unique solution.	

	But, never mind, suppose we somehow know that the coarse grained variables should
	be the environment in the   	
model of the previous section.  Suppose that (4.1) is the initial statevector. Let it
evolve for time $T'$ according to the Schr\"odinger 
evolution (3.6), at which time an experiment is done to see if the system is in state 
 $|\phi >\equiv \mu |a>+\nu |b>$ or  $|\phi' >\equiv \nu ^{*} |a>-\mu^{*} |b>$. The question 
we now ask is whether DHI assigns an event ($a$ or $b$) to a time $T<T'$.

	The answer to this, we are told, is to consider the magnitude 
	of the off-diagonal elements of the 
``Decoherence Functional," such as

$$\eqalign{&|{\rm Tr}\bigl\{e^{i\int_{T}^{T'}dtAW(t)}(|\phi ><\phi|e^{-i\int_{T}^{T'}dtAW(t)}
[|a><a|\rho (T)|b><b|]\bigr\}|=\cr
&\qquad\qquad\qquad\qquad\qquad\qquad
|\mu\nu\alpha\beta|e^{-{\lambda T'\over 2}(a-b)^{2}}\cr}\eqno(4.3)$$

\noindent (the decaying exponential is just $<{\rm env}_{b},T'|{\rm env}_{a},T'>$), 
where the projection operators $|a><a|$ and $|b><b|$ are applied at time $T$.
If these elements are small enough to be neglected 
(according to some prearranged limit),  an event is assigned to time $T$.

	But, as can be seen from Eq.(4.3), the off-diagonal elements do not depend upon $T$, but only 
upon $T'$.  That is, whether or not an 
 event takes place at an {\it earlier} time depends upon the time that the {\it later}  
experiment is performed. In DHI, the past depends upon the future.   
Here is an example of how the arbitrarily 
distant future can affect the past. Suppose the above evolution takes place
for an arbitrarily long time interval $T'$, after which  
the effectively time-reversed evolution
described in section 4.1 takes place for an interval $T'$, followed by
 the experiment at time $2T'$.  Due to the evolution between $T'$ and $2T'$, the
environment states coupled to $|a>$ and $|b>$ grow identical instead of near orthogonal,
and so the decoherence condition is not satisfied.
 Thus, because of this future 
evolution (between $T'$ and $2T'$), an event which would have otherwise occurred in 
the past (at $T<<T'$) never occurs at all.  

	The decoherence condition is designed so 
	NOWEN that DHI is wrong.
	If at any time in the future an experiment takes place 
	that measures interference between earlier states
	which would otherwise be said to represent earlier alternative events, DHI disallows
	the occurrence of these events. One could 
	imagine a theory in which the future has an effect on the immediate past, or a miniscule
	effect on the arbitrarily distant past, but 
it is hard to consider as satisfactory a description of past events which can be undone
 by  behavior in the arbitrarily distant future.
	It is only when the statevector evolution is all over
that DHI says for sure  what the histories are.
 If the world goes on forever, DHI
 may well be characterized by the New York Lottery slogan, ``Hey, ya never {\it know}."
 To have to wait until the end of the universe, or forever, 
 to be able to describe events is not a solution of the events problem. 

 	I shall conclude this section by mentioning the view a proponent of CSL could take of ESI
 	and DHI.  According to CSL, True Collapse from a 
 	superposition of states to one or another state is brought about by 
 	differences in particle locations in the different states: 
 	the more the differences, the faster the collapse.   
 	A decoherence condition
 	 may not predict
 	events in situations when they do occur and vice versa. 
 	However, there are cases where the differences in particle locations 
are such that CSL says events occur,
 	and the choice of the environment or coarse graining is such that
 False Collapsers will also say that events occur. Although it is unlikely
 	that the latter will get such details as the 
 	time of events right, these differences may be unobservable. 
 	 Thus these approaches may be regarded, just as CI may,
 	 as FAPP$^{11}$ procedures for predicting that events will occur in some cases.  
\bigskip
\noindent{\large{5 Index Model}}
\bigskip 
 	There is more in heaven and earth\dots
\hfill William Shakespeare
\medskip
	There are two reasons why the False Collapse approaches fail to satisfactorily handle events.
	The first is that they hope to rely upon known objects in spacetime to provide the
	preferred basis, but this does not seem to be do-able in a well-defined way.  The second
	is that the states describing 
such objects can evolve from near-orthogonality (which is what is used
	as a criteria for events) back to nonorthogonality (which undermines the predictions
	of events).  
	
		It is perhaps interesting to see that the program of describing events by a unitarily 
		evolving statevector {\it could} go through 
		if the objects which determine the preferred basis
		are not the usual objects in spacetime, and if the evolution is such that their
		 orthogonality, once achieved, cannot be undone. 		 
		 Models illustrating this can be
		 made,$^{12-14}$ and I shall end this paper by giving one,
		 partially to serve as a bridge on which False Collapsers and True Collapsers may meet.
		 
		 Consider the initial statevector (4.1) evolving unitarily according to the 
		 Schr\"odinger equation
		 
		 $${d|\psi,t>\over dt}=-i2\lambda A\Pi (t)|\psi,t>, \qquad
		 [W(t), \Pi(t')]=i\delta (t-t')\eqno(5.1)$$
		 
		 The $W(t)$'s are to represent new and fundamental variables of Nature
(beyond the usual particles and fields), each of which Nature allows to interact with
the system (composed of the usual particles and fields) only for a brief 
interval dt, and never again. The 
resulting statevector is

$$ \eqalign{&|\psi ,T>=\Bigl\lbrace\int D_{T^{+}}^{\infty}w
e^{-{1\over 4\lambda}\int_{T^{+}}^{\infty}dt
w^{2}(t)}|\{w\}>_{T^{+}}^{\infty}\Bigr\rbrace\cr
 &\cdot\int D_{0}^{T}w|\{w\}>_{0}^{T}
\Bigl\lbrace\alpha|a> e^{-{1\over 4\lambda}\int_{0}^{T}dt
[w(t)-2\lambda a]^{2}}+\beta |b>e^{-{1\over 4\lambda}\int_{0}^{T}dt
[w(t)-2\lambda b]^{2}}\Bigr\rbrace\cr} 
 \eqno(5.2a)$$
 
 \noindent which may more transparently, if less precisely, be written as
 
 $$\eqalign{&|\psi ,T>=\Bigl\lbrace|{\hbox{\rm state of uninteracted\ }}w(t),\ T<t<\infty >\Bigr\rbrace\cr
 &\cdot\sum_{{\rm all\ }w(t),\ 0\leq t\leq T}|{\hbox{\rm state labeled by\ }} w(t),\ 0\leq t\leq T>
\Bigl\lbrace|\psi ,T>_{w}^{CSL}\Bigr\rbrace\cr} 
 \eqno(5.2b)$$
 
		 According to Eq. (5.2), the evolution correlates 
		 each CSL state (2.5) (which evolves under a w(t)) to an ``environment" state
		 (labeled by that w(t)). The statevector is the ``sum" of these correlated product states 
		 (with the correct CSL probability weighting). Each of these states in the  
		 ``sum" is a history. One history, chosen on-line ``by God" (utilizing SQT's 
		 transition probabilities as each state ``forks" into
		 its unique orthogonal alternatives)  
		  represents reality and the others represent
		 unrealized possibilities. There is a satisfactory solution to the preferred basis
		 problem: the preferred basis is {\it well defined} by the orthogonal 
		 ``environment" basis. There is a satisfactory solution 
		 to the events problem: the nonorthogonal system states (correlated to the orthogonal 
		 ``environment" basis states)
		 describe the macroscopic system undergoing {\it True
		  Collapse} to state $a$ or $b$.
		  
		   	Of course, this model is physically 
		   	completely equivalent to the True Collapse model of section 2. 
		   It suggests that density matrix's False Collapse
		   only makes sense if, actually, it corresponds to statevector's True Collapse.
\bigskip
\noindent{\large{Acknowledgments}}
\bigskip 
 	Is it they don't understand what they are doing, or you don't understand what they
 	are doing? 
\hfill Bryce DeWitt
\medskip
	Although my talk at this conference was mostly a 
	review of CSL, I did in passing express my opinion that
 the decoherence approach does not solve the problems of SQT. 
This elicited the remark quoted above. I decided to write this paper to provide
evidence at least with regard to the second question.  I therefore thank Bryce DeWitt
for stimulating this work.  I would also like to thank Euan Squires for valuable
conversations, and for the warm hospitality at Durham University where most of this work was done,
and Guido Bacciagaluppi, Jeremy Butterfield, Meir Hemmo and Adrian Kent for critical comments. 
I also wish to thank the Hughes Foundation for a grant.

\bigskip
\noindent{\large{References and Remarks}}
\bigskip

\noindent 1. E. Schrodinger, Die Naturwissenschaften {\bf 23}, 807, 823, 844 (1935).
 
\noindent 2. J. S. Bell in {\it Sixty-Two Years of Uncertainty},
edited by A. Miller (Plenum, New York 1990), p. 17.	

\noindent 3. In this paper I shall not discuss interpretations where the statevector
is taken to correspond to many-anything.  These have their own versions 
of the events and preferred basis problems discussed here. 
Also, in what follows I shall
disregard the minor but subtle issue of nonevents which should be treated as events.

\noindent 4. P. Pearle, Physical Review A{\bf 39}, 2277 (1989).

\noindent 5. G. C. Ghirardi, P. Pearle and A. Rimini, Physical Review A{\bf 42}, 78 (1990).

\noindent 6. For reviews, see G. C. Ghirardi and A. Rimini in {\it  Sixty-Two Years of Uncertainty},
 edited by A. Miller (Plenum, New York 1990); 
 G. C. Ghirardi and P. Pearle in {\it Proceedings of the 
 Philosophy of Science Association 1990, Volume 2}, 
 edited by A. Fine, M. Forbes and  L. Wessels 
 (PSA Association, Michigan 1992); 
 G. C. Ghirardi in {\it Quantum Chaos-Quantum Measurement}, 
edited by P. Cvitanovic et. al., (Kluwer, the Netherlands 1992), p. 305;
P. Pearle, to be published in the {\it Proceedings of 
 the Cornelius Lanczos International Centenary Conference} (1994).

\noindent 7. H. P. Stapp, Phys. Rev. A{\bf 46}, 6860 (1992) introduced the nomenclature 
``Heisenberg reduction" and ``Von Neumann reduction" for what I call here True Collapse
and False Collapse respectively.

\noindent 8. See the paper in this volume by W. H. Zurek.

\noindent 9. See the papers in this volume by M. Gell-Mann and J. Hartle.

\noindent 10. 2.  J. S. Bell in {\it Quantum Gravity 2}, 
eds. C. Isham, R. Penrose and D. Sciama  (Clarendon Press, Oxford 1981), p. 611 
and in {\it Speakable and unspeakable in quantum mechanics}, 
(Cambridge University Press, Cambridge 1987), p. 117. 

\noindent 11. FAPP is a useful acronym coined by John Bell$^{2}$
meaning ``For All Practical Purposes," i.e., useful but not fundamentally sound.

\noindent 12. This formalism was introduced by V. P. Belavkin, 
Physics Letters A{\bf 140}, 355 (1989), based upon the work of 
 R. L. Hudson and K. R. Parthasarathy, Comm. Math. Phys. {\bf 93}, 301 (1984), to
 model continuous measurements.

\noindent 13. L. Di\'osi, in {\it Quantum Chaos-Quantum Measurement}, 
edited by P. Cvitanovic et. al., (Kluwer, the Netherlands 1992), p. 299.

\noindent 14. P. Pearle, Physical Review A {\bf 48}, 913 (1993); in {\it Stochastic Evolution
of Quantum States in Open Systems and in Measurement Processes}, eds. L. Di\'osi and
B. Luk\'acs (World Scientific 1994), p. 79.

\bye